\documentclass[
 twocolumn, pre, 10pt, aps,floatfix,
]{revtex4-2}

\usepackage{graphicx}
\usepackage{tikz}
\usepackage{dcolumn}
\usepackage{bm}
\usepackage{amsmath}
\usepackage{amssymb}
\usepackage{amsthm}
\usepackage{epsfig}
\usepackage{epsfig}
\usepackage{color}
\usepackage[colorlinks=true,link color=blue]{hyperref}
\usepackage{caption}
\usepackage{setspace} % Required package for line spacing
\usepackage{subcaption}
\usepackage[margin=10pt,font=small,labelfont=bf,
labelsep=endash]{caption}

\begin{document}
%\doublespacing

\title{Scaling Signatures of Electrical Glassiness in A$_2$SmTaO$_6$ (A = Ba, Sr, Ca)}
\author{Saswata Halder$^{1,2}$}
\email{\textcolor{black}{Corresponding author: shalder@hyd.amity.edu}}
\author{Binita Ghosh$^3$}
\author{T. P. Sinha$^2$}
\affiliation{$^1$Amity School of Engineering and Technology, Amity University Hyderabad, Hyderabad 500108, Telangana, India}
\affiliation{$^2$St. Paul's Mission Cathedral College, 33/1, Raja Rammohan Roy  Road, Kolkata 700009, India}
\affiliation{$^3$Department of Physics, Bose Institute, 93/1, Acharya Prafulla Chandra Road, Kolkata 700009, India}

\date{\today}
\maketitle
\newpage

\maketitle

\section*{abstract}
Disorder plays an important role in materials science, influencing material behavior across different length scales. Imperfections such as vacancies, atomic substitutions, lattice distortions, and microstructural inhomogeneity disrupt the ideal periodicity thereby altering the physical properties. Analogous to spin-glass systems, electrical ‘glassiness’ arises when charge carriers confront disordered energy landscapes, leading to a broad range of relaxation times, especially in polycrystalline materials where dipoles experience competing exchange interactions. Complex impedance, permittivity, and electric modulus distill out separate resistive and capacitive effects, offering insights into how microstructural inhomogeneities affects conduction mechanism. In the investigated polycrystalline double perovskites A$_2$SmTaO$_6$ (A = Ba, Sr, Ca), with power law driven ac conductivity, the hopping and relaxation of carriers are affected by both grains and grain boundaries. Scaling of the ac conductivity and impedance response reveals correlations between the conduction and relaxation timescales. The inhomogeneities in local energy landscape of ‘frustrated’ dipoles restrict the ‘universality’ of conduction mechanism across the bulk length scale.

\maketitle

\section*{Keywords}    
Scaling; Electrical conductivity; Impedance spectroscopy; Frustration; Disorder

\section{\label{sec:level1}Introduction}

Double perovskite oxides have garnered decades of research attention due to their exceptional chemical flexibility and a wide spectrum of tunable physical properties, such as high k dielectric constants, spin glass behavior, solar energy harvesting, and supercapacitor applications ~\cite{zhou2012synthesis, tang2022half, sarma2000magnetoresistance, chen2019rare, lim2016insights, chen2019dielectric, leng2022double, sahoo2019exchange, kearins2021cluster, yin2019oxide, halder2023tailored, chen2025review}. In most of these functionalities, electrical conduction plays a pivotal role in dictating the overall performance of the materials. Consequently, studying their electrical conductivity provides crucial insight into the charge transport processes operating within these structurally complex systems. The ac conductivity, shaped by both grain and grain boundaries, reveals how structural ordering, defect densities, and microstructural characteristics collectively influence transport behavior. Typically, frequency-dependent conductivity consists of two main components: (i) frequency-independent dc conductivity that follows a temperature-dependent power law and (ii) a frequency-dependent ac component that emerges above a characteristic frequency, known as the hopping frequency ($\omega_H$) ~\cite{jonscher1972frequency,jonscher1981new}. Interestingly, a wide variety of disordered solids, such as ion-conducting glasses ~\cite{lee1992ac}, amorphous and polycrystalline semiconductors ~\cite{mansingh1980ac, morocho2022ac}, conducting polymers ~\cite{papathanassiou2007universal}, transition metal oxides ~\cite{das2018ac}, organic-inorganic hybrids ~\cite{jellibi2016experimental}, and single crystals ~\cite{fangary2025influence}, exhibit remarkably similar ac conductivity spectra. The common factor across these diverse systems is their intrinsic 'disorder' of a specific form, which governs the universal features of their ac response. The ac conductivity spectra in all of these materials can often be collapsed onto a single master curve, underscoring the importance of scaling analysis in revealing key aspects of conduction processes in functional materials ~\cite{dyre2000universality}.

The scaling behavior of conductivity spectra has been widely investigated in glasses and amorphous systems ~\cite{lee1992ac,mansingh1980ac}; however, studies on polycrystalline perovskite oxides remain relatively scarce. In polycrsytalline perovskites, the conduction and relaxation processes are influenced by a variety of factors, including structural order, defect density, and microstructural characteristics. In a previous study, Halder et al. demonstrated the applicability of the time–temperature superposition principle (TTSP) by scaling the ac conductivity spectra in double perovskite A$_2$HoRuO$_6$ (A = Ba, Sr, Ca) ~\cite{halder2017time}. The dc conductivity ($\sigma_{dc}$), hopping frequency ($\omega_H$), and the relaxation frequency ($\omega_m$) were found to be closely aligned, indicating that the onset of ac conduction coincides with the frequency of the relaxation maxima. The mathematical expression used for the conductivity scaling was given by;
\begin{eqnarray}
\frac{\sigma_{ac}}{\sigma_{dc}} = F(\frac{\omega}{\omega_s})
\label{equation2}
\end{eqnarray}
where the scaling factor F is independent of temperature and $\omega_s$ is the scaling parameter. The conductivity spectra were successfully scaled using $\omega_s$ = $\sigma_{dc}$T ~\cite{summerfield1985universal} and $\omega_s$ = $\omega_H$ ~\cite{ghosh2000scaling,ghosh2001conductivity}. Although the scaled spectra for the grain boundary region collapsed onto a single master curve, those corresponding to the grain interior did not exhibit universal scaling. This discrepancy arises from the distinct thermal activation behaviors of charge carriers in the grain and grain boundary domains. Furthermore, to understand the effect of structural disorder on carrier concentration, a composition-based investigation of TTSP was carried out in the complex perovskite Ba$_2$HoRu$_{1-\textit{x}}$Sb$_\textit{x}$O$_6$, where it was shown that the conductivity spectra of all individual materials collapsed onto a single master curve when the composition (\textit{x}) was added to the scaling function F($\frac{\omega\textit{x}}{\omega_s}$) ~\cite{halder2020exploring}. In polycrystalline perovskite oxides, the distribution of grains and grain boundaries are not uniform but follow a distribution law. Therefore, relaxation behavior in these materials does not have a single relaxation time, but a distribution of relaxation times that accounts for the difference in polarization of the dipoles within these different microstructures ~\cite{kao2004dielectric}. Therefore, to critically analyze the effect of disorder on the conduction mechanism, one must consider the scaling behavior in each of these thermally activated regions separately.

\begin{figure}
	\begin{subfigure}[b]{0.2555\textwidth}
		\includegraphics[width=\linewidth]{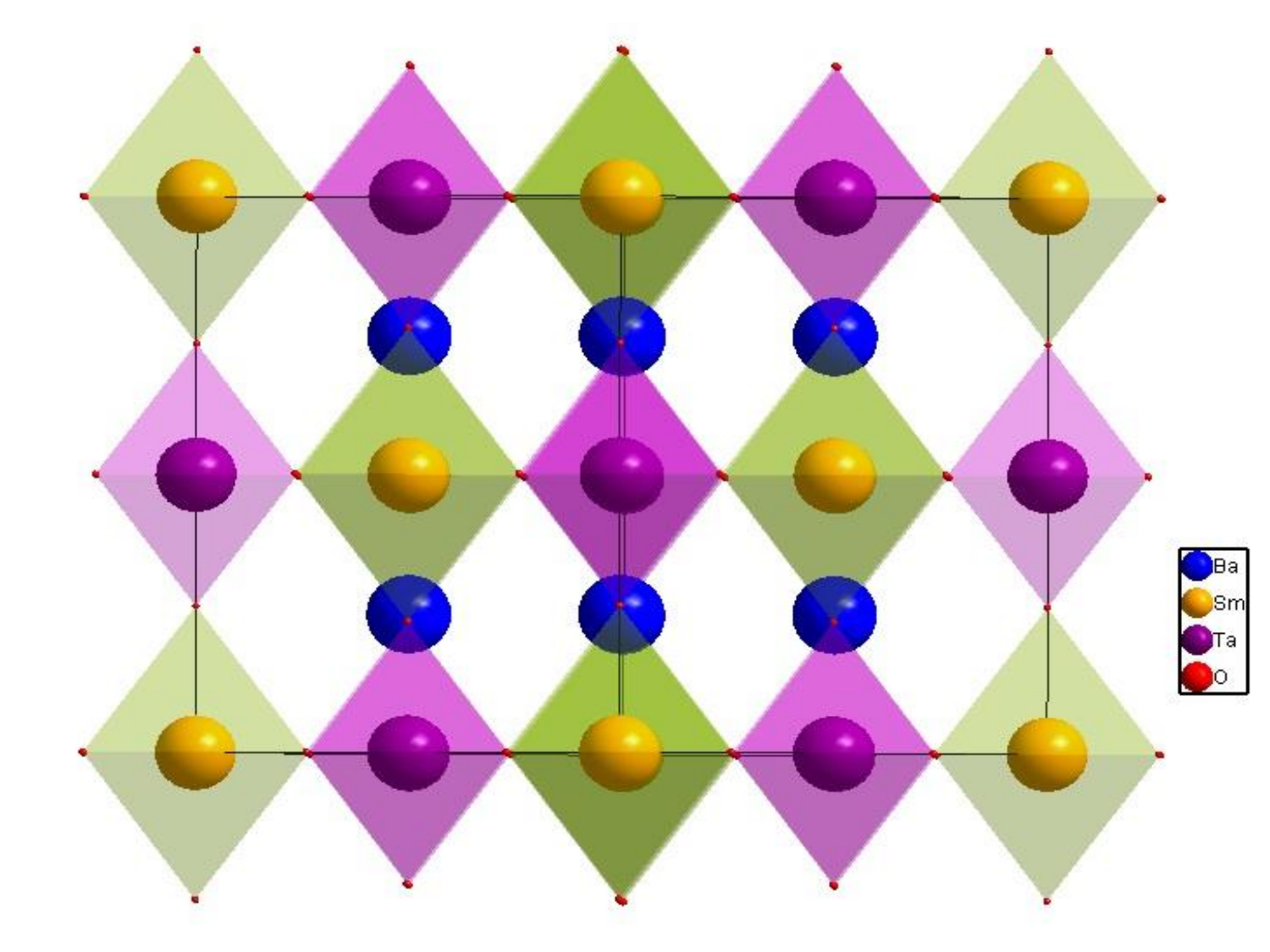}
		\caption{}
		\label{fig:crystalstructureBST}
	\end{subfigure}%
	\begin{subfigure}[b]{0.241\textwidth}
		\includegraphics[width=\linewidth]{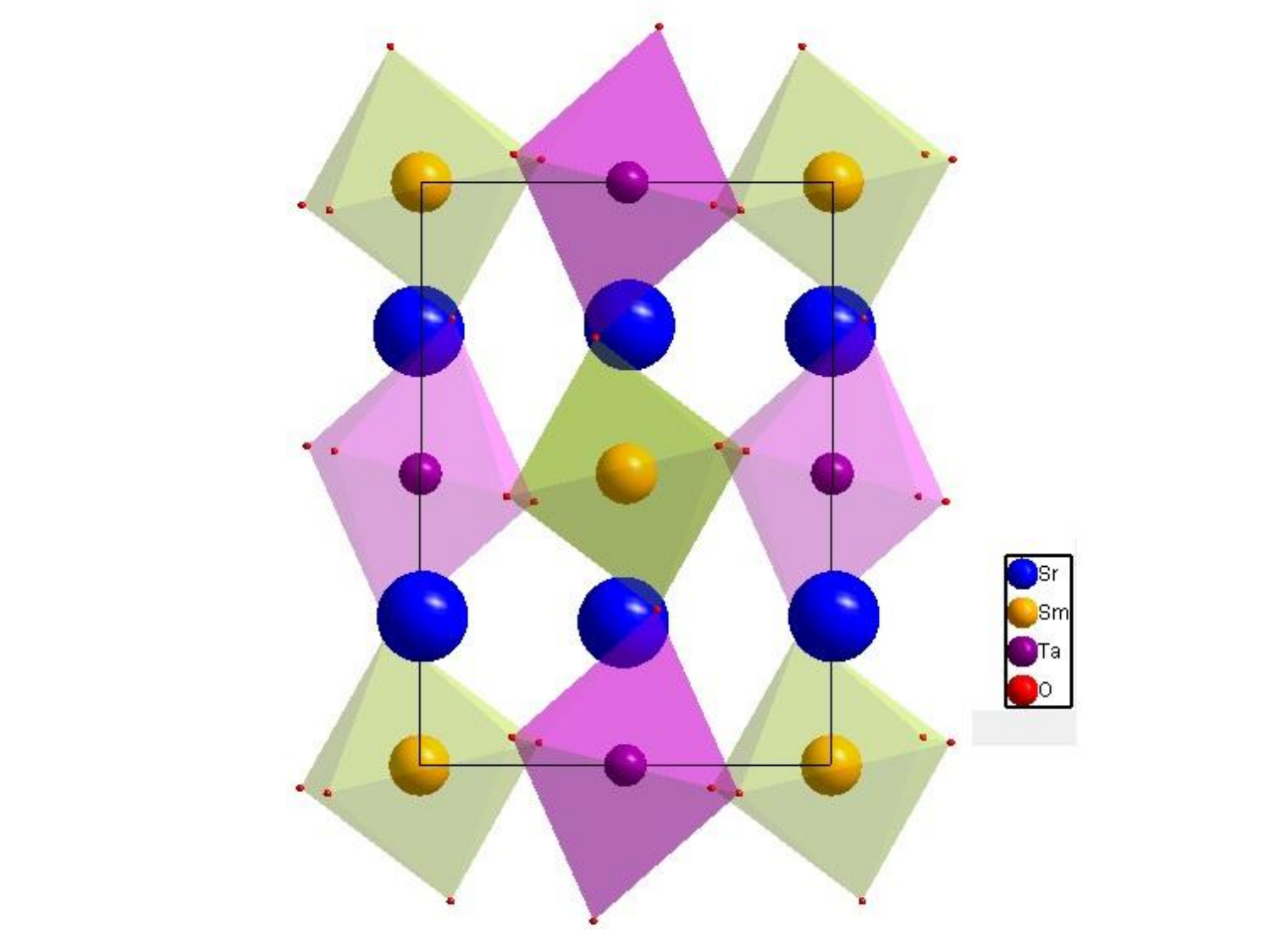}
		\caption{}
		\label{fig:crystalstructureSST}
	\end{subfigure}
       \begin{subfigure}[b]{0.2555\textwidth}
		\includegraphics[width=\linewidth]{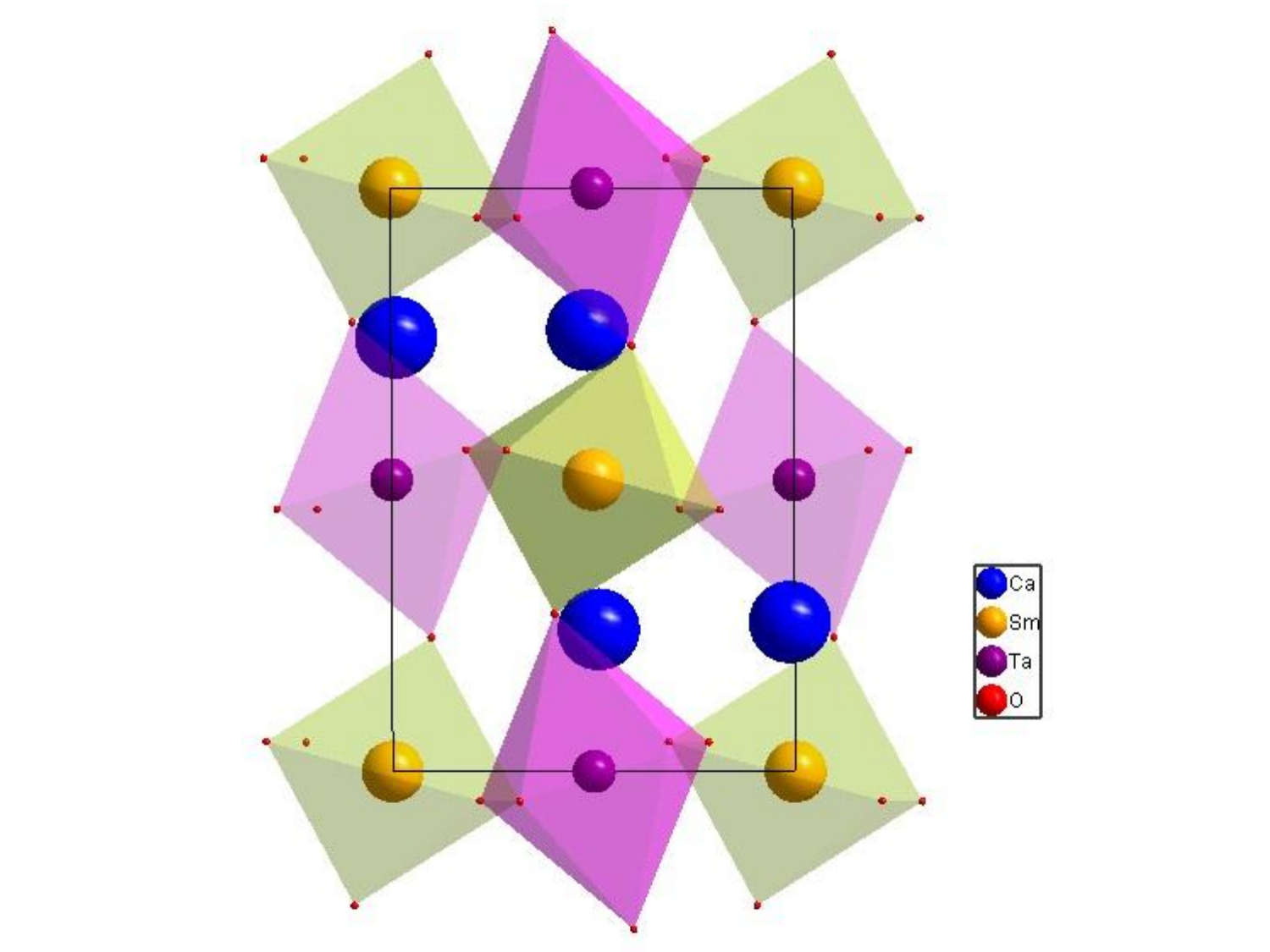}
		\caption{}
		\label{fig:crystalstructureCST}
	\end{subfigure}%
		\caption{Crystal structures for (a) BST, (b) SST, and (c) CST}
	\label{fig:crystalstructureAST}
\end{figure}

This article presents a comprehensive analysis of the dynamics of charge carriers in the 1:1 structurally ordered double perovskite oxides A$_2$SmTaO$_6$ (AST; A = Ba, Sr, Ca) at selected temperatures, using ac conductivity ($\sigma_{ac}$), electrical modulus (M) and electrical impedance (Z). In our earlier work, we explored the vibrational and conductivity properties of AST compounds in detail ~\cite{ghosh2014dielectric}. Ba$_2$SmTaO$_6$ (BST) was found to crystallize in a cubic \textit{Fm3m} structure, featuring a 1:1 ordering of Sm and Ta atoms at the B site, as depicted in Figure~\ref{fig:crystalstructureBST} \cite{ghosh2014dielectric}. This atomic ordering arises primarily because of differences in the ionic radii and electronegativities of the B-site cations. In contrast, Sr$_2$SmTaO$_6$ (SST) and Ca$_2$SmTaO$_6$ (CST) adopt a monoclinic structure with \textit{P2$_1$/n} symmetry (Figure~\ref{fig:crystalstructureSST} and ~\ref{fig:crystalstructureCST} respectively)~\cite{ghosh2014dielectric}. Like BST, SST and CST also exhibit 1:1 ordering of the SmO$_6$ and TaO$_6$ octahedra; however, octahedral tilting becomes evident due to the smaller ionic radii of Sr and Ca relative to Ba. This distortion increases from Sr to Ca. The electrical behavior of AST is strongly influenced by changes in crystal symmetry and the extent of hybridization between the Ta 5$\textit{d}$ and O 2$\textit{p}$ orbitals, which also leads to a decrease in the band gap, as seen in their electronic structure ~\cite{ghosh2014dielectric, ghosh2017x}. Unlike A$_2$HoRuO$_6$, the conductivity spectra of AST are predominantly governed by contributions from the grain boundary, especially at elevated temperatures. This makes them ideal candidates for investigating the scaling behavior of ac conductivity within the grain boundary regime. The observed Cole-Cole plots exhibit singular skewed semicircles, indicative of a distribution in relaxation times, often linked to nonuniform polarization processes within microstructural domains \cite{ghosh2014dielectric}. Consequently, AST double perovskites offer a promising platform to investigate the effects of structural disorder and heterogeneities in local microstructure on the carrier dynamics, due to the predominant contribution of the grain boundaries to the conduction mechanism. Understanding and controlling electrical disorder is vital for the design and optimization of polycrystalline oxides for a broad range of technological applications.

 \section{\label{sec:experimental}Experimental Details}
 
A detailed description of the synthesis of AST is provided in our previous communication ~\cite{ghosh2014dielectric}. The initial reactants BaCO$_3$, SrCO$_3$, CaCO$_3$, Sm$_2$O$_3$, and Ta$_2$O$_5$ are taken in stoichiometric ratios and mixed homogeneously by grinding in acetone medium for 10 h followed by calcination at 1370°C for 20 h. The powders were then allowed to cool to room temperature at a rate of 50°C/h. The sintering of the disk-shaped pellets (8 mm diameter and 1.25 mm thick) was carried out at 1390°C for 10 h followed by slow cooling to room temperature (40°C/h). The crystalline phase and lattice constants of the samples were determined by powder XRD (Rigaku MiniflexII diffractometer, Tokyo, Japan) using CuK$\alpha$ radiation. The structural and refinement parameters are provided in our previous communication ~\cite{ghosh2014dielectric}. The sintered pellets were polished, and electroded by high purity ultrafine silver paste for electrical characterization. The impedance (Z), phase angle ($\phi$) and conductance (G) of the samples were measured in a frequency window of 42 Hz to 5 MHz and at selected temperatures between 303 and 673 K using a computer-controlled LCR-meter (HIOKI-3552, Nagano,Japan). 

\section{\label{sec:level1}Results and Discussion}

\begin{figure}
	\begin{subfigure}[b]{0.511\textwidth}
		\includegraphics[width=\linewidth]{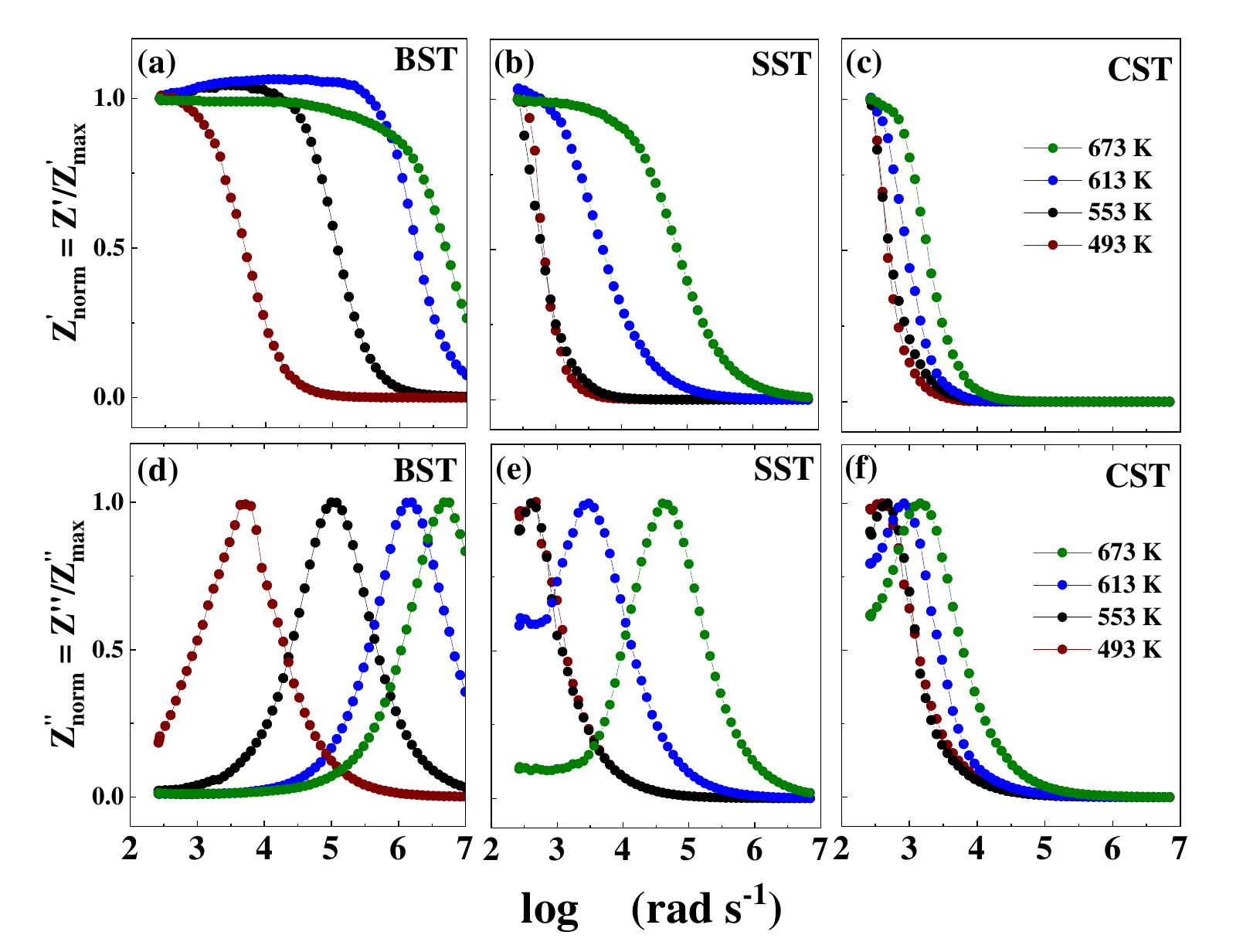}
	\end{subfigure}%
	\caption{Frequency-dependence of real and imaginary component of complex impedance Z* for (a), (d) BST, (b), (e) SST, and (c),(f) CST at selected temperatures.}
    \label{fig:ASTimpedance}
    \end{figure}

 Dynamical processes in materials are probed using a range of spectroscopic techniques, among which electrical relaxation measurements, typically frequency dependent, are widely employed to study charge dynamics. With advances in ac conductivity measurement methods, impedance spectroscopy has emerged as a powerful and reliable tool for analyzing experimental data. Figures ~\ref{fig:ASTimpedance}(a)–(c) present the real (Z$^{\prime}$) and imaginary (Z$^{\prime\prime}$) components of the complex impedance (Z$^*$)for BST, SST, and CST respectfully, at selected temperatures. The frequency-dependent behavior of the impedance (Z$^{\prime}$ reveals a shift in the relaxation response with increasing frequency, eventually reaching a plateau. This saturation at higher frequencies arises from the reduced polarization of the dipoles in response to the alternating electric field. A distinct peak in the imaginary part of the impedance Z$^{\prime\prime}$, centered around the relaxation region of Z$^{\prime}$, signifies the maxima of the relaxation process. As the temperature increases, this peak shifts toward higher frequencies, indicating a thermally activated relaxation phenomenon. In general, impedance spectra primarily capture contributions from the most resistive components in the system, such as grain boundaries. In contrast, grain contributions, which are more capacitive in nature, are better represented using the electric modulus formalism ~\cite{halder2017electronic}. As the frequency increases toward the peak, enhanced charge carrier mobility indicates long-range conduction. Beyond the peak, conduction is dominated by localized hopping, where carriers are trapped and move through short-range forward–backward hopping due to reduced mobility. Thus, $\omega_m$ represents the frequency limit for the transition from dc to dispersive conduction. The rise in temperature improves the dynamics of the charge carriers, decreasing the relaxation time ($\tau_m$; $\omega\tau$ = 1) and causing a corresponding shift of the relaxation frequency ($\omega_m$) to higher values. Interestingly, as we progress from BST to SST and CST, $\omega_m$ shifts toward lower frequencies, highlighting the increasing influence of high-resistance grain boundaries and slower relaxation dynamics.

%\textcolor{red}{Figures~\ref{fig:BSTconductivity(a),~\ref{fig:SSTconductivity}(a), and~\ref{fig:CSTconductivity}(a) show the ac conductivity plots for BST, SST, and CST, respectively, at selected temperatures. The frequency-dependent conductivity can be interpreted in terms of the grain boundary conductivity and jump relaxation model ~\cite{jonscher1972frequency,jonscher1981new}.The conductivity spectra follow the power law behavior, typical of semiconductors, given by:}

\begin{figure}
	\begin{subfigure}[b]{0.511\textwidth}
		\includegraphics[width=\linewidth]{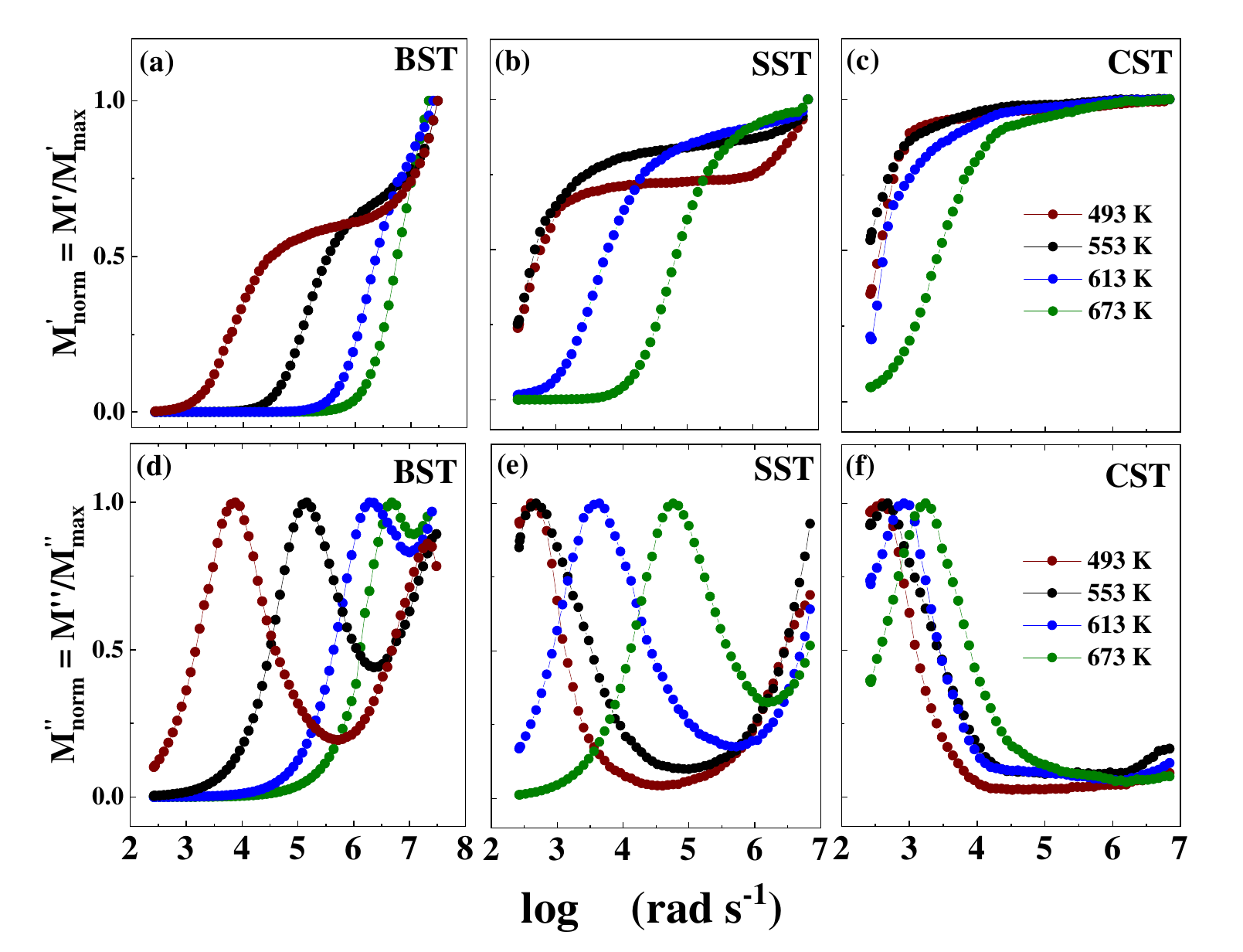}
	\end{subfigure}%
	\caption{Frequency dependence of complex electric modulus (M*) at selected temperatures. (a)-(c) represents the real component M'(w) for BST, SST and CST respectively. (d)-(f) represents the imaginary component M''(w) for BST, SST and CST respectively.}
    \label{fig:ASTmodulus}
\end{figure}

To complement the impedance analysis, the electric modulus formalism ($M^* = M^\prime + iM^{\prime\prime}$) was employed to investigate the dynamical aspects of electrical transport from a capacitive perspective ~\cite{halder2017electronic}. Unlike impedance spectroscopy, which is dominated by the response of highly resistive grain boundaries, the modulus representation emphasizes the relaxation processes associated with the conductive grain interiors. Thus, the combined analysis of the impedance and electric modulus spectra enables the decoding of the individual contributions of grains and grain boundaries. The frequency dependence of $M^\prime$ and $M^{\prime\prime}$ for BST, SST, and CST is shown in Figure~\ref{fig:ASTmodulus}. The real part, $M^\prime$, increases with frequency and approaches a constant value at higher frequencies due to the inability of dipoles to follow the rapidly varying electric field. The corresponding $M^{\prime\prime}$ spectra exhibit well-defined relaxation peaks, which confirm the dipolar relaxation mechanism in the material. With increasing temperature, the relaxation peaks shift toward higher frequencies, indicating thermally activated relaxation and a reduction in the relaxation timescale, similar to the $Z^{\prime\prime}$ spectra. At low temperatures, BST and SST exhibit two distinct relaxation features in both $M^\prime$ and $M^{\prime\prime}$, reflecting the simultaneous contributions of grains and grain boundaries. In contrast, for CST, the high-frequency relaxation peak shifts beyond the measured frequency window, indicating the dominant contribution of grain boundaries within the observable frequency window. At elevated temperatures, only a single relaxation peak is observed for all compositions in the observable frequency window, suggesting that grain boundary relaxation governs the electrical response. Similarly to impedance analysis, $\omega_m$ corresponds to a transition from long-range charge carrier motion to localized carrier hopping at defect sites. The relaxation time ($\tau_m$) is determined from $\omega_m$ using the relation $\omega_m\tau_m = 1$, while the broad and asymmetric nature of $M^{\prime\prime}$ indicates a distribution of relaxation times, characteristic of a non-Debye relaxation behavior.

\begin{figure}
	\begin{subfigure}[b]{0.511\textwidth}
		\includegraphics[width=\linewidth]{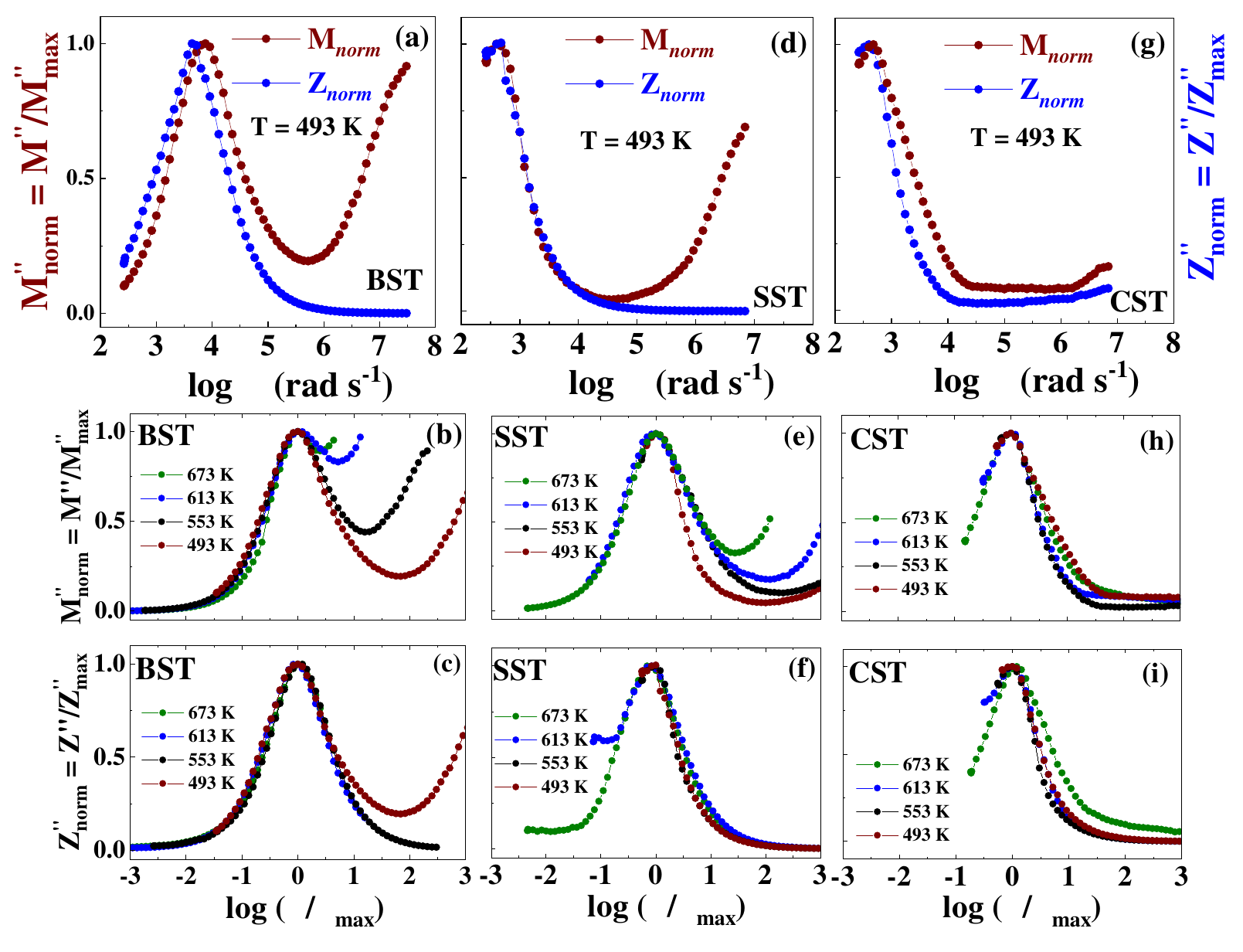}
		\end{subfigure}%
	\caption{Combined normalized electrical impedance-electrical modulus plots at 493 K for (a) BST, (d) SST and (g) CST. Scaling of the imaginary component of complex electric modulus (M*) for (b) BST, (e) SST, and (h) CST. Scaling of the imaginary component of complex electrical impedance (Z*) for (c) BST, (f) SST, and (i) CST.}
	\label{fig:ASTimpedancemodulus}
\end{figure}

The combined formalisms of impedance ($Z^*$) and electric modulus ($M^*$) provide complementary information on the electrical relaxation and conduction mechanisms by separating the resistive grain-boundary response from the bulk dipolar relaxation while minimizing the effects of electrode polarization ~\cite{halder2017electronic}. The relative positions of the imaginary impedance ($Z^{\prime\prime}$) and modulus ($M^{\prime\prime}$) peaks further distinguish long-range from localized charge transport. The coincidence of the $Z^{\prime\prime}$ and $M^{\prime\prime}$ peaks is characteristic of long-range charge carrier mobility with Debye-like relaxation, whereas their separation indicates that charge carriers remain spatially localized and relax through thermally activated hopping between energetically non-equivalent sites, giving rise to non-Debye behavior. Normalized Z$^{\prime\prime}$/Z$^{\prime\prime}_{max}$ and M$^{\prime\prime}$/M$^{\prime\prime}_{max}$ spectra for BST, SST, and CST are shown in Figure~\ref{fig:ASTimpedancemodulus}. As evident in Figures~\ref{fig:ASTimpedancemodulus}(a), \ref{fig:ASTimpedancemodulus}(d), and \ref{fig:ASTimpedancemodulus}(g), the maxima of $Z^{\prime\prime}$ occur at frequencies lower than the maxima of $M^{\prime\prime}$, confirming the relaxation is dominated by localized hopping rather than translational long-range conduction. At 493 K, both spectra are governed by a grain-boundary relaxation peak, while the high-frequency shoulder in the modulus spectra of BST, and SST signifies the emergence of the grain (bulk) response. Moreover, the broad and asymmetric relaxation peaks indicate a wide distribution of relaxation times, implying that charge carriers experience a broad spectrum of activation energies associated with structural disorder and defect-induced potential fluctuations. Such a hierarchy of energy barriers prevents the system from relaxing through a single characteristic time and instead produces cooperative, spatially heterogeneous dynamics over multiple time scales. This behavior is a hallmark of electrically glassy systems, where the freezing of charge carrier dynamics results from disorder-induced frustration.

The scaling behavior of the imaginary components of the electric modulus ($M^{\prime\prime}$) and impedance ($Z^{\prime\prime}$) was analyzed to investigate the relaxation dynamics in the AST compounds. The normalized spectra of $M^{\prime\prime}$ and $Z^{\prime\prime}$ for BST, SST, and CST are shown in Figures~\ref{fig:ASTimpedancemodulus}(b)–(c), (e)–(f), and (h)–(i), respectively. BST and SST exhibit good scaling in the grain boundary region, where the spectra at different temperatures collapse onto a single master curve at the grain boundary, indicating a universal relaxation mechanism governed by a relatively narrow distribution of relaxation times. The deviation in the high-frequency grain-interior region reflects microstructural heterogeneity and a slightly different energy activated relaxation mechanism but suggests weak electrical glassiness. In contrast, CST shows a breakdown of the scaling beyond $\omega_m$, indicating a wider distribution of activation energy barriers and relaxation times. The failure of M$^{\prime\prime}$ scaling in CST indicates that the relaxation-time distribution is temperature dependent. The charge-carrier dynamics is not governed by a single thermally activated relaxation process and is consistent with correlated and heterogeneous charge dynamics expected in an electrically glassy state. Therefore, the non-universal relaxation predicts an increase in the electrical glassiness for CST, resulting from a stronger disorder and spatially heterogeneous charge-carrier hopping compared to BST and SST.

\begin{figure}
	\begin{subfigure}[b]{0.511\textwidth}
		\includegraphics[width=\linewidth]{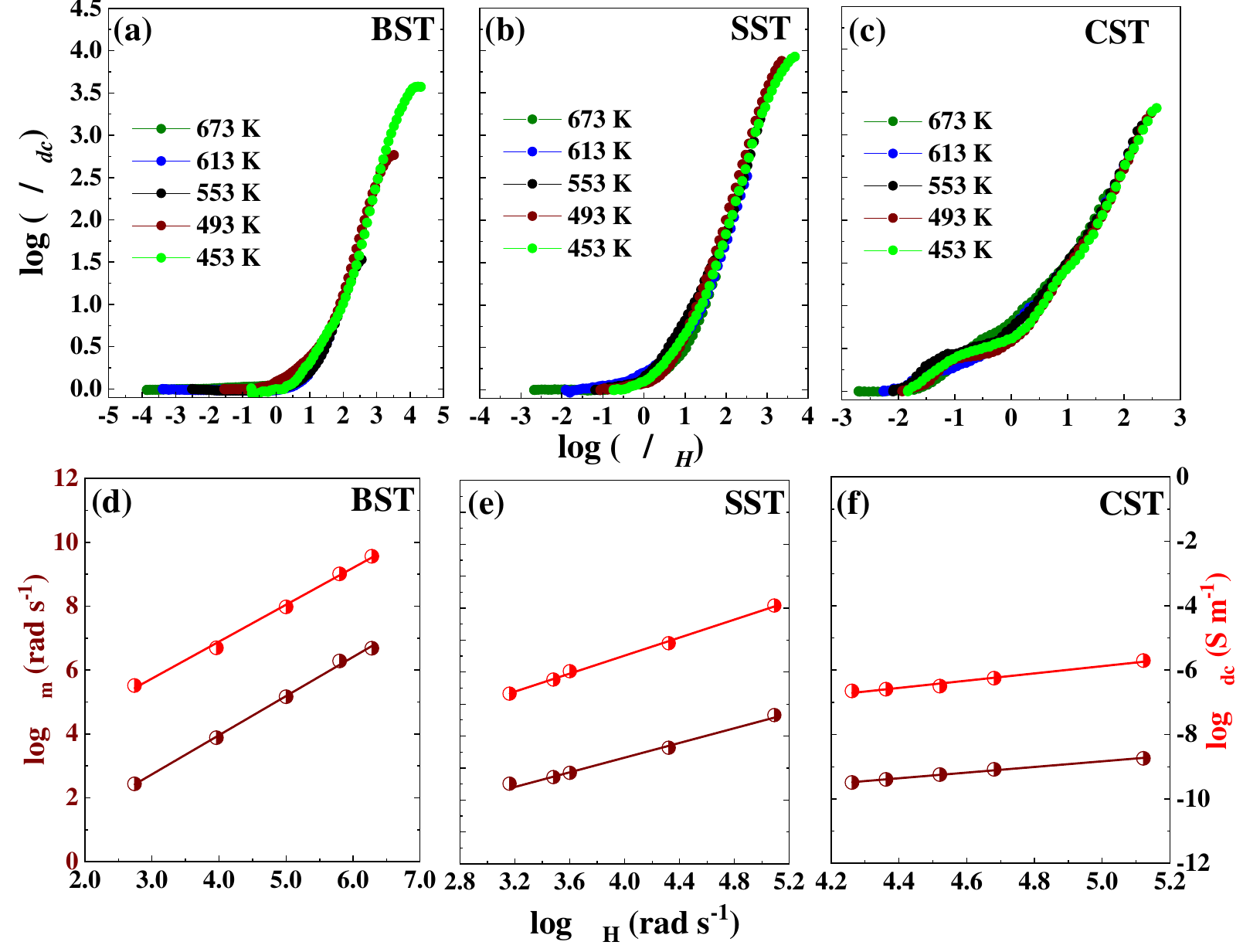}
		\end{subfigure}%
	\caption{Scaling of ac conductivity ($\sigma$$_{ac}$) for (a) BST, (b) SST, and (c) CST. Correlation between dc conductivity ($\sigma$$_{dc}$), relaxation frequency ($\omega$$_m$), and hopping frequency ($\omega$$_H$) for (d) BST, (e) SST and (f) CST.}
	\label{fig:ASTscaledconductivity}
\end{figure}

To gain a deeper understanding of the temperature dependence of the conduction mechanism and the distinct responses of grains and grain boundaries, the scaling of the frequency-dependent conductivity spectra has been analyzed. In this study, the frequency axis was scaled using $\omega_s = \omega_H$, while the conductivity axis was normalized by the dc conductivity ($\sigma_{dc}$) ~\cite{ghosh2000scaling,ghosh2001conductivity}. For BST, as shown in Figure ~\ref{fig:ASTscaledconductivity}(a), the scaling results in a complete collapse of the spectra to a single master curve, indicating a consistent scaling behavior. SST (Figure~\ref{fig:ASTscaledconductivity}b) also exhibit similar trends in the scaled conductivity spectra. A prominent deviation in the scaling of the conductivity spectra is observed for CST. The scaled conductivity spectra for CST (Figure ~\ref{fig:ASTscaledconductivity}c) do not collapse into a single master curve. Figures ~\ref{fig:ASTscaledconductivity}(d)-(f) indicate that the dc conductivity ($\sigma_{dc}$) and
hopping frequency ($\omega_H$) are strongly correlated as seen in glassy materials with a slope close to unity. It is important to note that deviations from ideal scaling emerge when disorder occurs within a similar microstructural environment. Since the grain contribution in BST and SST predominantly lies outside the frequency window, the minimal inhomogeneity mainly arises from the unequal response of dipoles trapped in the grain boundaries with different activation energies. For CST, this frequency window above $\omega_H$ has contributions to the conductivity from two different sources: (i) grain boundary relaxation and (ii) long range translation of carriers in grains. The disagreement in the scaling arises from the differently activated thermal transport in two different microstructural regions, which dominates the scaling deviations arising from inhomogeneities in the grain boundaries. 

To understand the electrical "glassiness" observed in AST, one can draw an analogy to the spin-glass behavior found in polycrystalline magnetic materials~\cite{sampathkumaran2007magnetic, gupta2019observation}. A spin-glass state arises from competing magnetic interactions that prevent the establishment of a true magnetic order, causing spins to freeze in random orientations. The associated energy landscape is highly complex, characterized by numerous local minima separated by energy barriers. These minima represent metastable configurations that do not necessarily correspond to the global energy minimum of a true magnetic ground state. The height and distribution of these barriers determine the system’s ability to explore the energy landscape, leading to slow non-equilibrium polarization dynamics. This frustrated landscape, resulting from random and competing interactions, results in a broad distribution of relaxation times, which is both temperature dependent and lacks a single characteristic timescale~\cite{sampathkumaran2007magnetic, murani1981spectral}. As temperature increases, the distribution of relaxation typically shifts towards shorter times (i.e. higher frequencies), reflecting faster dynamics. In ac susceptibility measurements, this behavior manifests itself as a frequency-dependent cusp in the real part of the susceptibility ($\chi^\prime$) at the spin-freezing temperature ($T_f$), while the imaginary part ($\chi^{\prime\prime}$) shows a peak at $T_f$ that shifts with frequency, a signature of spin-glass dynamics, distinguishing it from conventional magnetic ordering. A similar mechanism underlies the conduction and relaxation dynamics in perovskite oxides under an applied ac electric field. Here, charge carriers experience energy barriers associated with grain boundaries, which arise from a heterogeneous distribution of grains. As observed in Figure~\ref{fig:ASTimpedance}(a)-(c), the peak frequency ($\omega_m$) in the imaginary part of the impedance ($Z^{\prime\prime}$) shifts to lower frequencies from BST to CST, indicating slower relaxation. BST has a smoother energy landscape and faster charge dynamics, while CST exhibits slower relaxation processes. Analogously to spin frustration caused by competing exchange interactions, electrical relaxation in these oxides is influenced by local electrostatic environments that introduce a 'dipolar frustration.' This results in a wide range of relaxation times and sluggish charge dynamics. Furthermore, the trapping of charge carriers at the grain boundaries further contributes to the slow dynamics. The degree of this inhomogeneity and disorder in the energy landscape can be inferred from the scaling behavior of the conductivity and impedance spectra. A universal master curve typically indicates reduced disorder and more homogeneous dynamics, whereas deviations suggest that a higher degree of disorder impedes carrier transport.

\section{\label{sec:level1} Conclusions}

A quantitative analysis of the scaling behavior of ac conductivity ($\sigma_{ac}$), complex electrical modulus ($M^*$) and complex impedance ($Z^*$) has been carried out for polycrystalline double perovskite oxides A$_2$SmTaO$_6$ (AST; A = Ba, Sr, Ca) over a frequency range of 42 Hz to 5 MHz at selected temperatures. AST shows a 1:1 structural ordering at the B site, therefore eliminating any long-range disorder effects on the conduction process. The conductivity spectra primarily reflect contributions from grain boundary in the observed frequency window, with minor signatures from grains. A clear correlation is observed between the dc conductivity ($\sigma_{dc}$), the hopping frequency ($\omega_H$), and the relaxation frequency ($\omega_m$) indicating that the onset of dispersive conduction is intrinsically related to the relaxation dynamics of the system. The ac impedance spectra reveal a thermally activated relaxation mechanism characterized by a broad distribution of relaxation times. Among the studied samples, CST exhibits significantly slower relaxation dynamics compared to BST as seen from the longer relaxation times, which is an indicator of higher levels of electrical "frustration" due to competing energy barriers. Importantly, scaling analysis of the conductivity spectra reveals that the time–temperature superposition principle holds predominantly in the grain boundary regime, but even there, deviations from ideal behavior are evident, in the case of CST, due to local inhomogeneities in the grain boundary energy barriers. These findings highlight that the 'universal' nature of scaling behavior serves as a sensitive probe of the degree of disorder in the charge transport pathways. The extent to which conductivity spectra collapse onto a master curve reflects microscopic energetic inhomogeneity of the system, providing evidence for dipolar frustration and 'glassy' charge carrier hopping in these materials.

\section{\label{sec:level1} Declaration of Competing Interest}

The authors declare no competing financial conflicts of interest.

\bibliography{paper}
\bibliographystyle{unsrt}

\end{document}